\documentclass[
reprint,
superscriptaddress,
amsmath,
amssymb,
aps,
prl,
twocolumn,
footinbib]{revtex4-2}
\usepackage{graphicx}
\usepackage{dcolumn}
\usepackage{bm}
\usepackage{physics}
\usepackage{appendix}
\usepackage[percent]{overpic}
\usepackage[normalem]{ulem}
\usepackage{comment}
\usepackage{svg}
\usepackage{natbib}
\usepackage{todonotes}
\usepackage{csquotes}
\usepackage{algorithm}
\floatname{algorithm}{Protocol}
\usepackage{algorithmic}
\usepackage{bbm}
\usepackage{dsfont}
\usepackage{hyperref}
\usepackage{orcidlink}
\usepackage[capitalise]{cleveref}
\usepackage{placeins}

\usepackage{xcolor}

\newcommand{\ignore}[1]{}
\newcommand{\rd}{\mathrm{d}}
\newcommand{\rs}{\mathrm{s}}
\newcommand{\rx}{\mathrm{x}}
\newcommand{\re}{\mathrm{e}}
\newcommand{\cfi}{\mathpzc{F}}
\newcommand{\qfi}[1]{\mathcal{Q}#1}
\DeclareFontFamily{OT1}{pzc}{}
\DeclareFontShape{OT1}{pzc}{m}{it}%
              {<-> s * [1.25] pzcmi7t}{}
\DeclareMathAlphabet{\mathpzc}{OT1}{pzc}%
                                 {m}{it}

\begin{document}

\title{Exploring topological signatures of quantum Fisher information scaling in graph states}

\author{Majid Hassani\,\orcidlink{0000-0002-4077-0769}}
\email{majidhasani2010@gmail.com}
\affiliation{Instituut-Lorentz, Universiteit Leiden, P.O. Box 9506, 2300 RA Leiden, The Netherlands}
\affiliation{$\langle aQa^{L}\rangle$ Applied Quantum Algorithms Leiden, The Netherlands}
\author{Patrick Emonts\,\orcidlink{0000-0002-7274-4071}}
\affiliation{Instituut-Lorentz, Universiteit Leiden, P.O. Box 9506, 2300 RA Leiden, The Netherlands}
\affiliation{$\langle aQa^{L}\rangle$ Applied Quantum Algorithms Leiden, The Netherlands}
\affiliation{Institute for Complex Quantum Systems, Ulm University, 89069 Ulm, Germany}
\affiliation{Center for Integrated Quantum Science and Technology (IQST), Ulm-Stuttgart, Germany}
\author{Jordi Tura\,\orcidlink{0000-0002-6123-1422}}
\email{tura@lorentz.leidenuniv.nl}
\affiliation{Instituut-Lorentz, Universiteit Leiden, P.O. Box 9506, 2300 RA Leiden, The Netherlands}
\affiliation{$\langle aQa^{L}\rangle$ Applied Quantum Algorithms Leiden, The Netherlands}
\date{\today}
\begin{abstract}
Graph states are a versatile resource for quantum metrology, and their quantum Fisher information (QFI) has a closed-form expression in terms of the underlying graph: it counts pairs of vertices with identical neighborhoods. 
It remains unclear, however, which global structural properties of a graph determine whether its QFI exceeds the standard quantum limit (SQL). 
Here we ask whether topological invariants, namely the Betti numbers of the graph's clique complex, capture this behavior, and we address the question numerically with two complementary approaches. 
First, we randomly add or delete edges in balanced complete multipartite graphs, which combine a QFI beyond the SQL with a single nonvanishing higher Betti number. 
Edge addition eventually destroys both the metrological advantage and the nontrivial topology. 
Under edge deletion, by contrast, higher Betti numbers persist well after the QFI has dropped to the SQL. 
Second, we start from Erd\H{o}s--R\'enyi random graphs, whose QFI does not typically exceed the SQL, and maximize a chosen Betti number with a Monte Carlo search. 
The search drives the graphs toward near-extremal multipartite structures whose QFI lies well beyond the SQL. 
Our results show that topological optimization offers a route to metrologically useful graph states, but also that Betti numbers alone do not determine the QFI.
\end{abstract}

\maketitle
\section{Introduction\label{sec:introduction}}
The quantum Fisher information (QFI) plays a central role across a wide range of areas in quantum science. 
It acts as a powerful witness of multipartite entanglement~\cite{pezze_entanglement_2009,hyllus_fisher_2012,toth_multipartite_2012,strobel_fisher_2014,bohnet_quantum_2016}, quantifies information scrambling in chaotic quantum dynamics~\cite{wimberger_nonlinear_2014,shi_quantum_2025}, and serves as a central figure of merit in quantum metrology and sensing, underpinning both single-parameter~\cite{caves_quantum-mechanical_1981,pezze_quantum_2018,hassani2025many} and multiparameter estimation theory~\cite{ragy_compatibility_2016,demkowicz-dobrzanski_multi-parameter_2020,reilly_optimal_2023,annabestani_multiparameter_2022-1}. 
Moreover, it provides a quantitative measure of quantum macroscopicity~\cite{frowis_macroscopic_2018}. 
Beyond these applications, the QFI defines the Bures metric on the manifold of quantum states~\cite{sommers_bures_2003}.
A variety of its mathematical properties have been extensively studied, including convexity~\cite{fujiwara_fibre_2008,toth_extremal_2013,alipour_extended_2015-1}, continuity~\cite{augusiak_asymptotic_2016-1,rezakhani_continuity_2019-1}, and its geometrical interpretation~\cite{imai_semi-classical_2025}. 
Despite this widespread use and detailed mathematical understanding, identifying which structural features of many-body quantum states ultimately govern the scaling behavior of the QFI remains an open and actively investigated problem~\cite{pezze_quantum_2018,braun_quantum-enhanced_2018,degen_quantum_2017}.

A particularly natural setting to address this concern is provided by graph states~\cite{hein2006entanglement}. 
Graph states form a broad and well-studied class of multipartite entangled states, with applications ranging from computation~\cite{raussendorf_one-way_2001,nielsen_cluster-state_2006}, measurement based quantum computing~\cite{raussendorf_measurement-based_2003}, verification~\cite{markham_simple_2020}, and cryptography~\cite{markham_graph_2008,qian_quantum-cryptography_2012}, to quantum sensing~\cite{shettell_graph_2020,wang_continuous-variable_2020,bag2022achieving,tao2023quantum}. 
Their defining feature is that the entanglement structure of the state is entirely encoded in an underlying graph, whose vertices represent qubits and whose edges specify entangling operators. 
This explicit correspondence between quantum correlations and graph structure makes graph states a versatile platform for exploring nonclassical properties of quantum states.

While the scaling behavior of the QFI for graph states has been investigated previously~\cite{shettell_graph_2020}, its deeper connection to the topological properties of the underlying graphs has not yet been elucidated. 
In particular, existing analyses primarily rely on graph-theoretic descriptors that capture local or spectral features, leaving open the question of whether global topological characteristics play a role in determining the scaling of the QFI. 
Tools from algebraic topology provide a natural framework to address this gap. 
Within this framework, Betti numbers emerge as fundamental topological invariants that quantify the number of independent connected components, cycles, and higher-dimensional cavities present in a graph or, more generally, in an associated simplicial complex.

Recently, Betti numbers have emerged as central descriptors in topological data analysis (TDA), where they are widely used to characterize the global structure of complex data sets beyond purely geometric or statistical features~\cite{moroni_learning_2021,zhang_persistent_2025}. 
By quantifying the number of connected components, cycles, and higher-dimensional topological features across different scales, Betti numbers provide a compact and robust summary of topological complexity that has proven valuable in fields ranging from TDA~\cite{gyurik_towards_2022} to quantum many-body physics~\cite{sehayek_persistent_2022,olsthoorn_persistent_2023}. 
At the same time, the classical computation of Betti numbers for large and highly connected systems is known to be computationally demanding, motivating the development of quantum algorithms capable of achieving polynomial or even exponential speedups in their estimation~\cite{lloyd_quantum_2016,berry_analyzing_2024}. 

Since the metric aspects of the QFI have already been extensively investigated, here we instead focus on its relation to topological characteristics of multipartite quantum states. To this end, we employ graph states as the central framework of our analysis. In Ref.~\cite{shettell_graph_2020}, a closed analytical expression for the QFI of graph states was derived, providing a direct connection between graph structure and metrological performance. At the same time, the topology of the underlying graph can be characterized through its Betti numbers, which quantify nontrivial higher-order connectivity properties. Motivated by these observations, in this work we investigate \textit{whether topological quantities can serve as a descriptor for the scaling behavior of the QFI with the total number of qubits $N$}. In particular, we study the relationship between the scaling of the QFI, which determines the attainable quantum-enhanced precision, and the Betti numbers associated with graph states defined on $N$ qubits.

While the QFI admits an analytical formula for arbitrary graph states, analytical expressions for the Betti numbers are only available for restricted graph families, such as balanced complete $l$-partite graphs. Consequently, our investigation is primarily based on numerical analysis. We approach the problem from two complementary directions.

First, we consider balanced complete $l$-partite graphs, which simultaneously exhibit super-linear QFI scaling and nontrivial topology. We then perturb these graphs through random edge addition and random edge deletion processes. In the case of edge addition, increasing the perturbation probability rapidly drives the QFI toward linear scaling, thereby eliminating the metrological advantage. Simultaneously, the higher-order Betti numbers quickly vanish, indicating that the topology of the underlying graph becomes progressively trivial. In contrast, under edge deletion the QFI again approaches linear scaling; however, the higher-order Betti numbers display greater robustness. Their decay is substantially slower than in the edge-addition scenario, and for sufficiently strong perturbations the graph can still retain nontrivial topological features even when the QFI exhibits only linear scaling.

Second, we investigate ensembles of random graphs with a fixed number of qubits. We find that uniformly sampled random graphs typically do not provide a metrological advantage, as their QFI scales linearly with system size. However, when the graphs are iteratively modified through an optimization procedure that maximizes a selected higher-order Betti number $\beta_k$, the resulting graph families develop enhanced metrological properties. More specifically, once the bundled structure forms, the QFI grows beyond the linear scaling regime as the targeted Betti number increases, while at convergence the targeted Betti number approaches its maximal value and the remaining Betti numbers vanish. These observations suggest that higher-order topological invariants can capture structural features relevant to the emergence of quantum-enhanced metrological scaling.

Altogether, our results indicate that Betti numbers, although purely topological quantities, can act as an informative descriptor for the scaling behavior of the QFI in graph states. 
This provides evidence for a previously unexplored connection between topological graph invariants and multipartite quantum metrology, and may provide new insights into the design of highly sensitive quantum states through topological principles.

This paper is organized as follows. In Sec.~\ref{Prelimi}, we briefly review the necessary preliminaries of quantum estimation theory. Since our work focuses on the relationship between the QFI of graph states and their Betti numbers, we first introduce the basic concepts of QFI in Sec.~\ref{Intro_QMetrolog}. In Sec.~\ref{Intro_qfi_graph_states}, we then review graph states and present the canonical expression for their QFI. Finally, in Sec.~\ref{Intro_QFI_metric}, we examine the connection between the QFI and a class of well-known metrics on the manifold of quantum states.
Independently, in Sec.~\ref{Intro_Betti}, we introduce Betti numbers from the perspective of TDA. 
Our main results on the relationship between the QFI and the Betti numbers of graph states are presented in Sec.~\ref{Sec_QFI_Betti}, where we analyze perturbations of the balanced complete $l$-partite graph in Sec.~\ref{Pertur_Bundled_Graph} and investigate optimal configurations within random graphs in Sec.~\ref{Opt_Rand_Graph}. 
Finally, Sec.~\ref{Sec_conclusion} concludes the paper with a summary and outlook.

\section{Preliminaries}\label{Prelimi}
In this section, we present the preliminaries and necessary background for our results.
\subsection{Single parameter quantum metrology}\label{Intro_QMetrolog}
Quantum metrology and sensing constitute a central pillar of quantum technologies, exploiting quantum correlations to enhance the estimation of unknown parameters in dynamical processes~\cite{giovannetti_quantum_2006,paris_quantum_2009,giovannetti_advances_2011,escher_general_2011,degen_quantum_2017,hassani_digital_2017,pezze_quantum_2018,polino_photonic_2020,cassens_entanglement-enhanced_2025}. 
A typical protocol proceeds in three stages. 
First, a probe state $\rho_{0}$ is prepared (preparation stage). 
Next, the probe interacts with the dynamics that encode the parameter of interest $\theta$ (sampling stage). 
Finally, a measurement is performed using positive operator-valued measure (POVM) with elements $\{\Pi_{\rx}\}$ satisfying $\sum_{\rx}\Pi_{\rx}=\mathbf{1}$ (measurement stage), yielding the conditional probability distribution of outcomes $p(\rx|\theta)$~\cite{giovannetti_quantum_2006}. 
The fundamental bound on the attainable precision in parameter estimation is given by the Cram{\'er}–Rao inequality~\cite{hayashi_asymptotic_2005}
\begin{equation}\label{classical_cramer_rao}
    \delta\theta \geqslant \frac{1}{\sqrt{\nu\cfi}},
\end{equation}
where $\delta\theta$ denotes the estimation error, defined as the standard deviation of an unbiased estimator, $\nu$ is the number of independent repetitions, and $\cfi$ is the classical Fisher information (CFI). 
The latter is defined as
\begin{equation}\label{def_CFI}
    \cfi = \sum_{\rx} p(\rx|\theta) \left[ \frac{\partial}{\partial\theta} \log p(\rx|\theta) \right]^{2},
\end{equation}
and quantifies the information about $\theta$ that can be extracted from the measurement outcomes $\{\Pi_{\rx}\}$. 
The CFI generally depends on the unknown parameter $\theta$, see \cref{def_CFI}.
To understand how the achievable precision bound scales with physical resources, it is instructive to consider probe states composed of multiple particles. 
In this setting, for a product state of $N$ particles, the measurement outcomes can be treated as independent and identically distributed (i.i.d.) random variables. 
Here, the CFI scales as $\mathcal{O}(N)$, which corresponds to the so-called standard quantum limit (SQL).

By applying the Born rule, $p(\rx|\theta)=\Tr[\Pi _{\rx}\rho _{\theta}]$ where $\rho _{\theta}$ denotes the probe state after the sampling stage, and substituting into \cref{def_CFI}, optimization of the CFI over \textit{all} possible measurements yields the QFI
\begin{equation}\label{def_QFI}
    \qfi(\rho _{\theta})=\Tr[\rho _{\theta}L_{\theta}^{2}],
\end{equation}
where $L_{\theta}$ is the symmetric logarithmic derivative (SLD), defined by
\begin{equation}
    \partial_{\theta} \rho _{\theta}=\frac{L_{\theta}\rho _{\theta}+\rho _{\theta}L_{\theta}}{2},
\end{equation}
in which $\partial _{\theta}:=\partial/\partial\theta$. 
The corresponding quantum Cram{\'e}r–Rao bound reads
\begin{equation}\label{quantum_cramer_rao}
    \delta\theta \geqslant \frac{1}{\sqrt{\nu\qfi(\rho _{\theta})}},
\end{equation}
which sets the ultimate lower limit on the achievable estimation error for an unbiased estimator~\cite{caves_quantum-mechanical_1981,braunstein_statistical_1994,braunstein_generalized_1996,paris_quantum_2009}. 
In quantum metrology, preparing the probe state $\rho_{0}$ as an entangled state of $N$ particles enables the QFI to scale as $\mathcal{O}(N^{2})$, corresponding to the Heisenberg limit (HL)~\cite{giovannetti_quantum_2006}. 
In the case of pure probe state $\rho _{0}=\ket{\psi}\bra{\psi}$ and unitary encoding dynamics, the SLD has the simple form of $L_{\theta}=2\partial_{\theta}\rho_{\theta}$. 
Hence the QFI is given by
\begin{equation}\label{QFI-pure-states}
  \qfi(\rho_{\theta})= 4\left(\braket{\partial_{\theta}\psi_{\theta}}{\partial_{\theta}\psi_{\theta}}-\vert\braket{\psi_{\theta}}{\partial_{\theta}\psi_{\theta}}\vert ^{2}\right).
\end{equation}
From the general formalism outlined above, the QFI can be evaluated for any parameterized family of quantum states. We now specialize this framework to the class of graph states, which are central to our analysis. 
\subsection{Quantum Fisher information of graph states}\label{Intro_qfi_graph_states}
Let $G = (V, E)$ be a simple graph, where $V=\{v_{1},\cdots,v_{N}\}$ denotes the set of vertices and $E \subseteq \{e_{i,j} \mid v_{i},v_{j} \in V, i \neq j\}$ the set of edges.

\begin{figure}[h]
    \centering
    \includegraphics[width=0.3\textwidth]{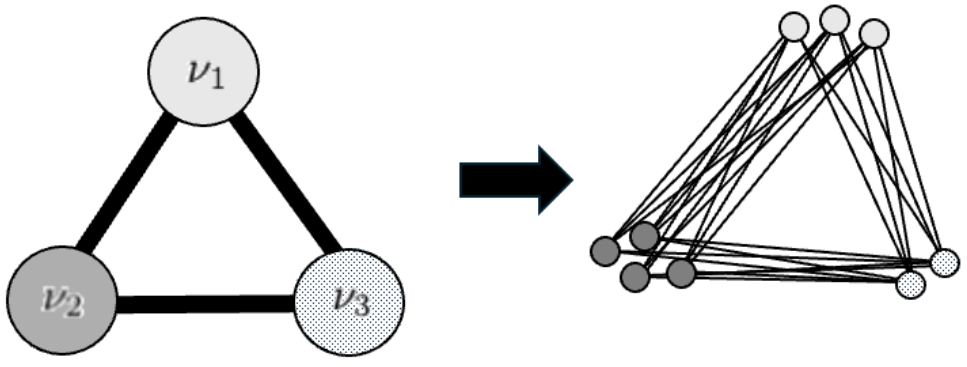}
    \caption{A bundled graph state with $N=9$ qubits and three islands with $\nu_{1}=3$, $\nu_{2}=4$, and $\nu_{3}=2$.}
    \label{bundled_graph_3}
\end{figure}

A quantum state can be associated with each graph in the form of a \textit{graph state}. 
To define this correspondence, each vertex $v_{i}\in V $ is identified with a qubit, while each edge $e_{i,j}\in E$, connecting two vertices $v_{i}$ and $v_{j}$, represents a quantum correlation implemented by an entangling operation. 
In this formalism, all qubits are initialized in the state $\ket{+}=1/\sqrt{2}(\ket{0}+\ket{1})$ and a controlled-$Z$, ($CZ$), $CZ_{(i,j)}=\ket{0}_{i}\bra{0}\otimes\mathbf{1}_{j}+\ket{1}_{i}\bra{1}\otimes Z_{j}$, is applied to each pair of qubits $(i,j)$ connected by an edge. 
Consequently, the graph $G=(V,E)$ defines the quantum state
\begin{equation}
    \ket{G}=\prod_{(i,j)\in E}CZ_{(i,j)}\ket{+}^{\otimes N}.
    \label{eq:def_graph_states}
\end{equation}
Notably, \cref{eq:def_graph_states} includes GHZ states, which have many applications in quantum metrology and networked sensing~\cite{shettell_graph_2020,shettell_private_2022,hassani_privacy_2025}. 
Now, the quantum state $\ket{G}$ can be used as an input probe state for a phase estimation problem, in which the parameter encoding is implemented by the unitary operator
\begin{equation}
    U_{\theta} =\text{Exp}\left[-i\frac{\theta}{2}\sum _{i=1}^{N}\sigma^{i} _{x}\right],
\end{equation}
where $\sigma^{i}_{x}$ denotes the Pauli-$X$ operator acting on the $i$th qubit. 
Following Ref.~\cite{shettell_graph_2020}, the QFI of the evolved graph state $\ket{G_{\theta}}=U_{\theta}\ket{G}$ can be computed using the stabilizer formalism as 
\begin{equation}\label{QFI_graph_1}
    \qfi(G_{\theta})=\sum _{i,j=1}^{N}\delta _{N(i),N(j)},
\end{equation}
where $\delta$ is the Kronecker delta, $N(i)$ represents the neighborhood of vertex $i$, and the graph has no isolated vertices, i.e., $\vert N(i)\vert\geqslant 1,~\forall i\in V$.
By partitioning the vertex set into disjoint subsets $\mathcal{V}_{1}, \mathcal{V}_{2}, \cdots, \mathcal{V}_{L}$ such that $N(i)=N(j)$ for all $i,j\in \mathcal{V}_{l}$ and $|\mathcal{V}_{l}|=\nu_l$, \cref{QFI_graph_1} reduces to
\begin{equation}\label{QFI_graph_2}
\qfi(G_{\theta})=\sum_{l}\nu_l^2.
\end{equation}
The sets $\{\mathcal{V}_{l}\}_{l=1}^{L}$ are precisely the equivalence classes of the relation $u\sim v$ if and only if $N(u)=N(v)$, where $L$ denotes the number of distinct neighborhood classes. If $G$ contains isolated vertices, each corresponding qubit is in the state $\ket{+}$, which is an eigenstate of the generator and therefore contributes zero to the QFI. Accordingly, Eq.~\eqref{QFI_graph_2} remains valid, with the sum restricted to equivalence classes with nonempty neighborhoods. Hence
\begin{equation}
\qfi(G_{\theta})=\sum_l \nu_l^2 \geqslant \frac{N^2}{L}.
\end{equation}
Therefore, the QFI directly quantifies the redundancy of vertex neighborhoods; it is enhanced when many vertices share identical neighborhoods and exhibits super-linear scaling whenever the number of distinct neighborhood classes satisfies $L=o(N)$. This observation naturally motivates the construction of \textit{bundled graph states}, introduced in Ref.~\cite{shettell_graph_2020},
which are designed to maximize such neighborhood redundancy. 
A bundled graph consists of several sets of vertices, called islands, where the qubits within each island are connected to all qubits in the other islands, but not to those within the same island. 
In other words, we consider bundled graphs constructed from the complete graph $K_l$, resulting in complete $l$-partite graphs. 
Figure~\ref{bundled_graph_3} illustrates a bundled graph with three islands and a total of $N=9$ qubits, where $\nu_1=3$, $\nu_2=4$, and $\nu_3=2$. 
From \cref{QFI_graph_2}, the QFI for this graph is
$\qfi = \sum_{i=1}^{3}\nu_i^2 = 29 \approx 9^{1.5}$. 
Note that this bundled graph state does not quite saturate the HL with an exponent of two, but exceeds the SQL.
We next consider the balanced case, in which each of the $l$ islands contains $m$ vertices, giving a total of $N=ml$ vertices (qubits). The resulting graphs are balanced complete $l$-partite graphs (Tur\'an graphs), which coincide with the bundled cyclic graphs of Ref.~\cite{shettell_graph_2020} for $l=3$. For this class of graphs, the QFI is given by
\begin{equation}\label{QFI_bundled_cyclic}
    \qfi(G_{\mathrm{balanced\ complete}})= \frac{N^{2}}{l}=m^{2}l .
\end{equation}


\subsection{Metric property of the QFI}\label{Intro_QFI_metric}

In addition, from a geometric perspective, the QFI is intimately related to the Fubini-Study metric, which quantifies the distinguishability between neighboring quantum states in the projective Hilbert space with normalized vectors. 
Consider two infinitesimally close pure states, $\ket{\psi_{1}}=\ket{\psi}$ and $\ket{\psi _{2}}=\ket{\psi+\rd\psi}$. 
The infinitesimal angle ($\rd\rs$) between them is defined as
\begin{equation}\label{infin_angle}
    \rd\rs=\cos ^{-1}\big(\lvert\braket{\psi_{1}}{\psi_{2}}\rvert\big).
\end{equation}
Expanding to second order in $\rd\psi$ yields
\begin{equation}\label{FS_metric}
    \rd\rs ^{2}= \braket{\rd\psi}{\rd\psi} - \abs{\braket{\psi}{\rd\psi}}^ {2},
\end{equation}
which defines the Fubini-Study metric~\cite{gibbons_typical_1992,schack_hypersensitivity_1994}. 
For pure states, the QFI is equal to four times the Fubini-Study metric, as expressed in \cref{QFI-pure-states}. 
This correspondence can be generalized to mixed states. 
Starting from \cref{infin_angle}, one can write $\cos(\rd\rs)=\abs{\braket{\psi_1}{\psi_2}}$ and expanding the cosine function to second order gives
\begin{equation}\label{B_metric_deriv}
    1-\frac{\rd\rs ^{2}}{2}=\abs{\braket{\psi_1}{\psi_2}}.
\end{equation}
The right-hand side of \cref{B_metric_deriv} can be generalized to mixed states by replacing the overlap with the fidelity, $F=\Tr[\sqrt{\sqrt{\rho}\sigma\sqrt{\rho}}]$, between arbitrary density matrices $\rho$ and $\sigma$. 
Equation~\eqref{B_metric_deriv} then takes the form 
\begin{equation}\label{B_metric}
    D^{2}(\sigma,\rho)= 2(1-F(\rho,\sigma)),
\end{equation}
where $D^{2}(\sigma,\rho)$ defines the Bures metric. 
Substituting $\sigma=\rho_{\theta}+\delta\rho_{\theta}$ into \cref{B_metric}, yields~\cite{sommers_bures_2003}
\begin{equation}
    \qfi(\rho _{\theta})=\frac{4D^{2}(\rho_{\theta}+\delta\rho_{\theta},\rho_{\theta})}{\delta\theta^{2}},
\end{equation}
demonstrating that the QFI is four times the metric on the manifold of parameterized quantum states. 
The metric nature of the QFI~\cite{imai_semi-classical_2025,wilson_geometric_2025} underlies its broad applicability to various subjects, including quantum phase transitions~\cite{campos_venuti_quantum_2007,zanardi_quantum_2008} and holography~\cite{lashkari_canonical_2016}. 
While the geometric aspects of the QFI and their physical implications have been extensively studied, here we establish a connection between the QFI and the topological properties of the underlying graph.
\subsection{Topological data analysis and Betti numbers}\label{Intro_Betti}
Having established how the QFI can be evaluated for graph states, we now turn to the structural properties encoded by the underlying graphs. Beyond their role in defining quantum states, graphs constructed from data naturally capture geometric information, which can be further analyzed from a topological perspective. In particular, TDA provides a systematic framework to extract robust scale-dependent features such as connected components, cycles, and higher-dimensional voids. To make this connection precise, we briefly review the main concepts of TDA. 

Extracting meaningful information from large and complex datasets is often a challenging task. 
TDA provides a geometric framework for this purpose by employing algebraic tools to identify and quantify invariant features of the dataset. 
In TDA, a dataset is represented as a collection of points embedded in a metric space. 
One option is to construct a graph in which the vertices ($V$) correspond to data points and edges ($E$) connect pairs of points that are \textit{close} to each other. 
To quantify this \textit{closeness}, we introduce a distance measure $d_{ij}$ between two arbitrary points $i$ and $j$. 
An edge is drawn between them whenever $d_{ij}\leqslant\varepsilon$, where $\varepsilon$ denotes the grouping scale~\cite{edelsbrunner_computational_2010,carlsson_topology_2009,adamaszek_extremal_2014,otter_roadmap_2017}.

After constructing the graph $G=(V,E)$, one can associate it with a particular type of simplicial complex known as a \textit{clique complex}, where a 1-clique corresponds to a vertex, a 2-clique to an edge, a 3-clique to a triangle, a 4-clique to a tetrahedron, and, in general, a $(k+1)$-clique to a $k$-dimensional simplex. 
While TDA is naturally formulated in terms of simplicial complexes built from data points, the graph used here is merely a convenient representation. 
To extract the topological properties of a dataset, it is crucial to identify its  $k$-dimensional holes, which can be computed from the homology groups using linear algebra on boundary operators. 
The boundary operator $\partial _{k}^{G}$ maps each ($k+1$)-clique to a formal linear combination of its constituent $k$-cliques. Its adjoint, $(\partial_k^G)^\dagger$, defines the corresponding co-boundary operator. From these operators, the \textit{combinatorial Laplacian} can be defined as follows
\begin{equation}\label{comb_Lap}
    \Delta _{k}^{G}=(\partial _{k}^{G})^{\dagger}\partial _{k}^{G}+ \partial _{k+1}^{G} (\partial _{k+1}^{G})^{\dagger},
\end{equation}
where the dimension of its kernel yields the $k$-th Betti number
\begin{equation}
    \beta _{k}^{G}=\text{dim ker}(\Delta _{k}^{G}),
\end{equation}
which quantifies the number of $k$-dimensional holes in the associated clique complex, i.e., $(\beta _{0},\beta _{1},\beta _{2},\cdots\beta _{k})$ correspond to the numbers of connected components, loops, voids, and their higher-dimensional analogs in the dataset~\cite{lloyd_quantum_2016,gyurik_towards_2022,berry_analyzing_2024}. The emergence of nonzero higher Betti numbers indicates nontrivial topology.

As an illustrative example of a clique complex exhibiting a large Betti number, we consider balanced complete $l$-partite graphs (with $l$ islands, each containing $m$ inhabitants). For this class of graphs, the $(l-1)$th Betti number is given by~\cite{berry_analyzing_2024}
\begin{equation}\label{Betti_bundled_cyclic}
    \beta _{l-1}=(m-1)^{l}.
\end{equation}
Notably, this is the extremal value, which maximizes $\beta_{l-1}$ over all graphs on $N$ vertices~\cite{kozlov_convex_1997,beers2025extremal}.
\section{The QFI and Betti numbers in graph states}\label{Sec_QFI_Betti}
\begin{figure*}[t]
    \centering
    \includegraphics[width=0.87\textwidth]{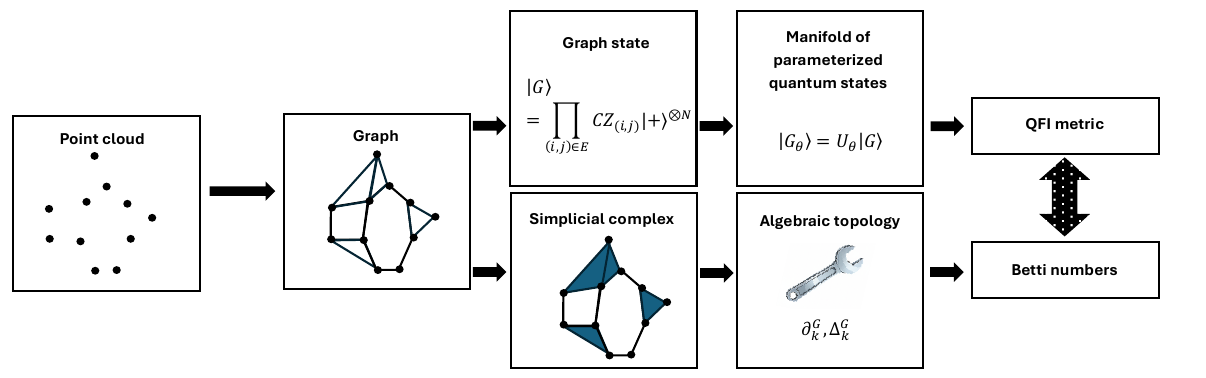}
    \caption{Schematic overview of the paper. 
    A dataset is represented as a point cloud in an embedded space. 
    By introducing a grouping scale $\varepsilon$, points that are sufficiently close are connected, forming a graph representation of the data. 
    From this graph, one can construct simplicial complexes and, using tools from algebraic topology, compute the corresponding Betti numbers, which quantify the $k$-dimensional holes in the dataset. 
    Alternatively, a quantum graph state can be associated with the same graph, where the vertices represent qubits and the edges correspond to $CZ$ operations. 
    Under parameter-encoding unitary evolution, this graph state maps onto a manifold of parameterized quantum states, on which the QFI defines a natural metric. 
    The objective of this paper is to explore the relationship between the QFI and the Betti numbers of the underlying graph structure.}
    \label{sch_paper}
\end{figure*}

As discussed in \cref{Intro_Betti}, a dataset can be represented as a point cloud in an embedding space. From such a representation, one can construct a simplicial complex that encodes relations between data points. TDA then proceeds by computing invariants such as Betti numbers using tools from algebraic topology. In this work, we focus on the graph representation of the data together with its associated simplicial structure, which form the basis for defining the corresponding graph states --- see the lower branch in Fig.~\ref{sch_paper}.

In parallel, once the graph is constructed, a corresponding quantum graph state can be associated with it, where the vertices represent qubits and the edges correspond to $CZ$ operators. 
This graph state can then undergo parameter-encoding unitary evolution, mapping it onto a manifold of parameterized quantum states. 
As discussed above, the QFI provides a natural metric on this manifold --- see the upper branch in Fig.~\ref{sch_paper}. 
Given the existence of a closed-form expression for the QFI of graph states, \cref{QFI_graph_2}, we can systematically evaluate the QFI across different graph instances. 

A natural question then arises as to whether there exists a relationship between the QFI and the topological characteristics of the underlying graph, such as the Betti numbers. Establishing such a connection would provide a bridge between quantum information-geometric quantities and topological invariants of the underlying structure. To investigate this possibility, we perform numerical analyses and summarize our findings below. A schematic overview of the present study is shown in Fig.~\ref{sch_paper}.

We \textit{numerically} consider two complementary approaches. 
First, we start from balanced complete $l$-partite graph states (as ideal reference cases) that simultaneously exhibit a metrological advantage through QFI scaling beyond the SQL, \cref{QFI_bundled_cyclic}, and nontrivial topology characterized by a nonzero $k$-th Betti number, \cref{Betti_bundled_cyclic}. We then perturb these graph states through random edge addition or deletion.

Second, using a Monte Carlo search algorithm to add or remove individual edges, we start from a random graph and attempt to maximize one of the Betti numbers. 
When the Monte Carlo converges, it finds the appropriate bundled graph. 
During this optimization, as the target Betti number increases, the QFI again scales beyond the SQL while the other Betti numbers vanish. 
As a technical caveat, the closed-form expression for the QFI of a graph state in \cref{QFI_graph_1} is valid \textit{only} for graphs with no isolated vertices. Whenever a graph with $\beta_0>1$ occurs in our numerics, we assign $\mathcal{Q}=N$ by convention. Therefore, we restrict ourselves to graphs satisfying $\beta_0=1$, a sufficient, though not necessary, condition for having no isolated vertices (with the exception of the trivial single-vertex graph).

Throughout this work, Betti numbers are computed using the persistent homology algorithm implemented in the Python package \texttt{Ripser.py}~\cite{bauer2021ripser}, applied to the clique complex of the underlying graph. 
To compute a target Betti number $\beta_k$, the simplicial complex is constructed up to dimension $k+1$.
\subsection{Perturbation on balanced complete $l$-partite graphs}\label{Pertur_Bundled_Graph}
To explore the relation between the QFI and the Betti numbers, and given the difficulty of obtaining an analytical connection, we adopt a numerical approach. 
We consider two balanced complete $l$-partite graphs, denoted by $\ket{G^{(\mathbf{3})}}$ and $\ket{G^{(\mathbf{4})}}$, characterized by $l=3$ and $l=4$ islands, respectively. 
In both cases, each island contains an equal number of inhabitants, $\nu_i=20$. 
The corresponding system sizes, QFI, and Betti numbers are computed by \cref{QFI_bundled_cyclic,Betti_bundled_cyclic} and summarized in \cref{tab:bundled_examples}.

Starting from these reference states, we introduce perturbations by adding or deleting edges in the underlying graphs. Specifically, in the edge addition protocol, each pair of vertices not connected in the reference graph is independently connected with probability $p_{\text{edge addition}}$, which we refer to as the edge addition probability. Conversely, in the edge deletion protocol, each existing edge is independently removed with probability $p_{\text{edge deletion}}$. At the level of graph states, these modifications are implemented by applying $CZ$ gates between selected pairs of qubits, which create or remove edges depending on the initial connectivity.

\begin{table}[h]
\centering
\caption{Graph parameters, QFI, \cref{QFI_bundled_cyclic}, and nonzero Betti numbers, \cref{Betti_bundled_cyclic}, for two balanced complete $l$-partite graph states, $l=3$ and $l=4$, with $m=20$.}
\label{tab:bundled_examples}
\small
\renewcommand{\arraystretch}{1.3}
\begin{tabular}{|c|c|c|c|}
\hline
\textbf{State} & $\boldsymbol{N}$ & $\qfi$ & \textbf{Nonzero Betti numbers} \\
\hline
\strut $G^{(\mathbf{3})}$ & 60 & 1200 & $\beta_0 = 1,\ \beta_2 = 6859$ \\
\hline
\strut $G^{(\mathbf{4})}$ & 80 & 1600 & $\beta_0 = 1,\ \beta_3 = 130321$ \\
\hline
\end{tabular}
\end{table}

\subsubsection{Random edge addition}
\begin{figure*} 
\includegraphics[width=1\linewidth]{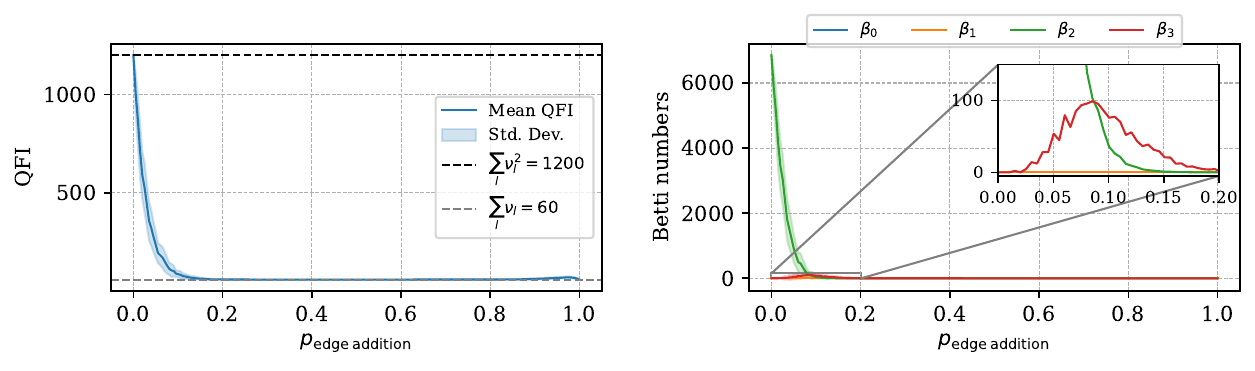}\quad\includegraphics[width=1\linewidth]{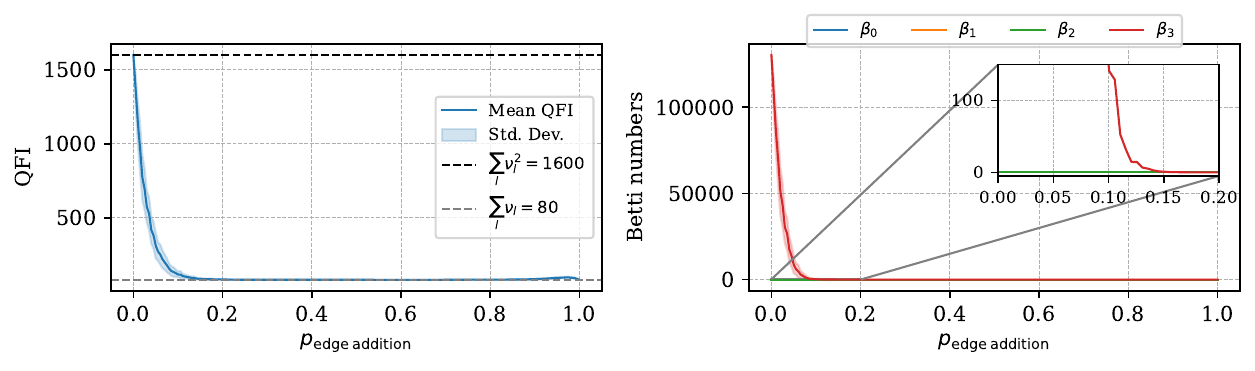}
\caption{Random edge addition to bundled graph states, averaged over $100$ realizations. 
\emph{Top}: three islands ($N=60$); \emph{bottom}: four islands ($N=80$), with $20$ nodes per island. 
\emph{Left panels}: QFI as a function of the edge addition probability $p$. 
\emph{Right panels}: Betti numbers as a function of $p$. 
Insets show that for $p > 0.2$, all Betti numbers vanish except $\beta_{0}$, indicating that only the connectivity of the graph remains.} 
\label{bundled_add_edge} 
\end{figure*}

We begin with the states $\ket{G^{(\mathbf{3})}}$ and $\ket{G^{(\mathbf{4})}}$ and iteratively add edges at random between previously unconnected vertices. 
For each bundled graph, this random edge addition protocol is repeated $100$ times. 
\Cref{bundled_add_edge} displays the resulting behavior of the QFI and the Betti numbers as functions of the edge addition probability for $\ket{G^{(\mathbf{3})}}$ and $\ket{G^{(\mathbf{4})}}$.
In both cases, under weak perturbations ($p\lesssim 0.1$), the value of the QFI remains in the beyond-SQL regime, and the corresponding graph states preserve nontrivial topology characterized by nonzero $\beta_{2}$ and $\beta_{3}$, respectively. Moreover, under edge addition, $\beta_{3}$ exhibits a peak for $0.02\lesssim p\lesssim 0.18$ in the three-island graph.

For $p> 0.2$, the value of the QFI is in the SQL regime and all Betti numbers, except $\beta _{0}$, vanish, see insets in right panels of Fig.~\ref{bundled_add_edge}. 
\subsubsection{Random edge deletion}
\begin{figure*} 
\includegraphics[width=1\linewidth]{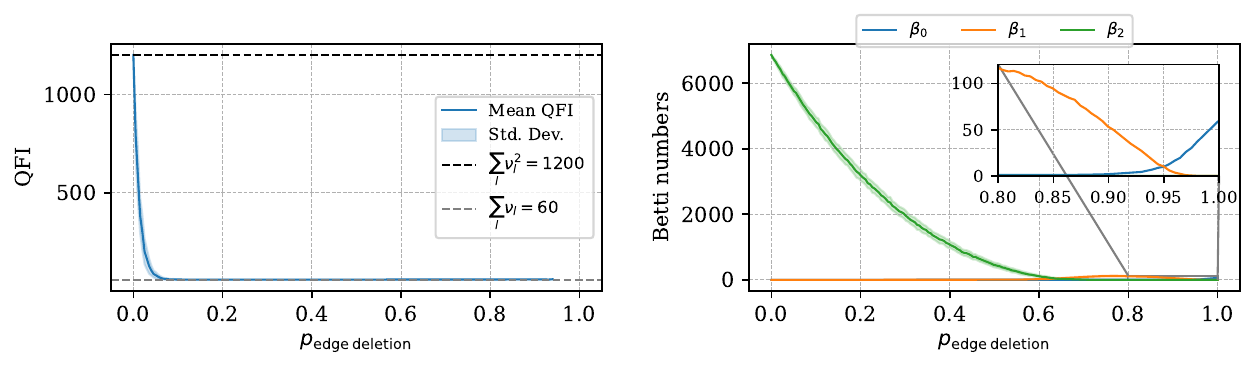}\quad\includegraphics[width=1\linewidth]{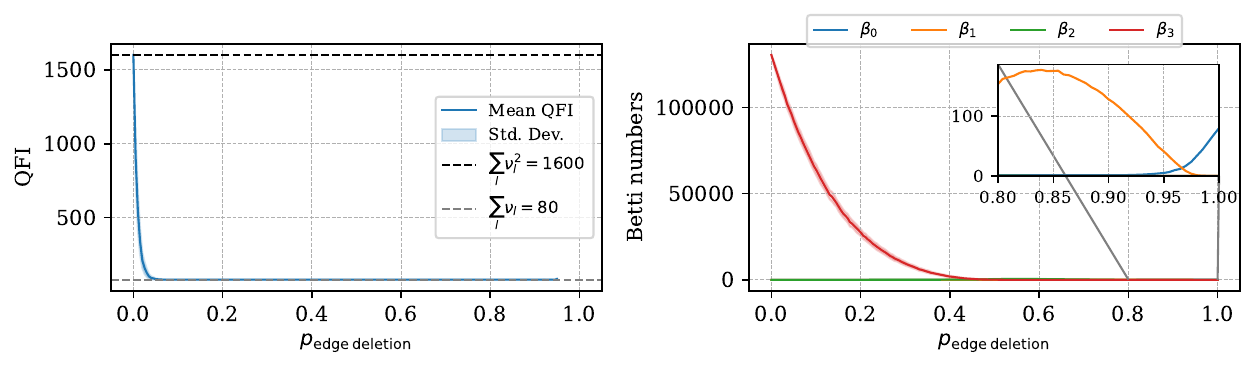}
\caption{Random edge deletion applied to bundled graph states, averaged over $100$ realizations. 
\emph{Top}: three islands ($N=60$); \emph{bottom}: four islands ($N=80$), with $20$ nodes per island. 
\emph{Left panels}: QFI as a function of the edge deletion probability $p$. 
\emph{Right panels}: Betti numbers as a function of $p$. 
Insets show that as $p \to 1$, all Betti numbers vanish except $\beta_{0}$, which approaches $60$ and $80$, respectively, indicating that all edges have been removed. In this regime, \cref{QFI_graph_2} is not applicable because isolated vertices appear, and $\qfi=N$ is used by convention for plotting purposes.}
\label{bundled_remove_edge} 
\end{figure*}

As an additional perturbation of the balanced complete $l$-partite graphs, we randomly delete edges between pairs of nodes in the graph states $\ket{G^{(\mathbf{3})}}$ and $\ket{G^{(\mathbf{4})}}$, performing $100$ repetitions for each state. 
The resulting behavior of the QFI and all Betti numbers is shown in Fig.~\ref{bundled_remove_edge} for $\ket{G^{(\mathbf{3})}}$ and $\ket{G^{(\mathbf{4})}}$, respectively. 
In both cases, the QFI surpasses the SQL regime only for very weak perturbations, corresponding to edge deletion probabilities of a few percent. 
In fact, the QFI already approaches the SQL at $p=0.1$, despite the persistence of nontrivial topological structure. 
For larger perturbation strengths ($p \geqslant 0.1$), the QFI falls within the SQL regime. 
In this regime, the topology of the graph can remain nontrivial, as reflected by the persistence of nonzero higher Betti numbers. 
Moreover, unlike balanced complete $l$-partite graphs, whose topology is characterized by a single nonzero higher Betti number ($\beta_2$ and $\beta_3$, respectively), the perturbed graphs exhibit multiple nonzero Betti numbers, indicating a more intricate topological structure. 

In contrast to the edge addition case, the highest Betti number is more robust under edge deletion than the QFI. 
This difference arises because deleting even a single edge can take a vertex out of its original equivalence class, thereby strongly affecting the metrological properties of the graph state, whereas the highest Betti number can decrease only gradually as edges are removed. 
Thus, the QFI is substantially more sensitive to weak edge deletion perturbations than the corresponding topological invariant.

We also note that bundled star graphs are topologically equivalent to bundled graphs with two islands, where all peripheral islands can be grouped into a single effective island connected to the central one. 
In this representation, the only nonvanishing Betti number is $\beta_1$. 
We have verified numerically that the results obtained above also apply to a two-island graph with $\nu_1=\nu_2=20$ (not shown), except for the case in which the central island contains \textit{solely one} vertex, which we discuss in the conclusion.

In the perturbed ensembles studied here, the loss of the metrological advantage goes together with the loss of nontrivial topology under edge addition, but not under edge deletion. 
However, the converse does not generally hold, as the intermediate-$p$ regime of Fig.~\ref{bundled_remove_edge} already illustrates.
\subsection{Optimizing random graphs}\label{Opt_Rand_Graph}
In the previous section, we observed that achieving QFI scaling beyond the SQL constrains and shapes the topological structure of the underlying graph in the weak perturbation regime. 
Here, we investigate the relationship in the opposite direction. 
Instead of analyzing how QFI and Betti numbers respond to perturbations, we attempt to construct graphs with a prescribed topological structure and then examine how the QFI of the corresponding graph states scales throughout this process. 
To this end, we begin from a random graph with a fixed number of vertices. 
By randomly adding or deleting edges, we iteratively modify the graph in order to maximize a chosen Betti number. 
During this optimization, we simultaneously track the evolution of the QFI of the associated graph state. 
The maximization of the selected Betti number is performed using a Metropolis Monte Carlo (MC) based search procedure~\cite{metropolis_equation_1953,liu_monte_2004,newman_monte_2010}. 
For brevity, we refer to this procedure as MC search throughout the remainder of the paper.
\subsubsection{MC search}
We employ an MC search to optimize the graph toward larger values of a chosen Betti number. 
Starting from an Erd\H{o}s--R\'enyi random graph $G$, we select a target Betti number $k$ and compute its initial Betti numbers. 
At each MC step, we propose a random modification of the graph by selecting two distinct vertices $u$ and $v$ uniformly at random and either adding or removing the edge $(u,v)$, depending on whether it is absent or present. 
This proposed graph $G'$ differs from $G$ by exactly one edge.

We then recompute the $k$-th Betti number and evaluate the change
\begin{equation}\label{delt_betti_k_metro}
    \Delta\beta _{k}=\beta _{k}(G')-\beta _{k}(G).
\end{equation}
The proposed update is accepted with the standard MC search probability
\begin{equation}\label{prob_metro}
    p = \min\left(1, \re^{\Delta\beta_{k} / T}\right),
\end{equation}
where $T$ is an effective temperature that controls the acceptance of moves with $\Delta\beta_{k}<0$. 
If the update is accepted, we set $G \leftarrow G'$, and update the stored Betti number; otherwise the graph remains unchanged. 
This procedure is iterated for a fixed number of steps, generating a trajectory of graphs whose Betti numbers and QFI values we record at each iteration, Protocol~\ref{Metropolis_algorith_format}.
We note that at low temperatures $T$, the acceptance probability $p$, \cref{prob_metro}, suppresses moves that decrease the target Betti number. 
When the protocol reaches a configuration where almost every possible edge flip would reduce $\beta _{k}$, the MC search becomes effectively stuck in a local maximum: proposed updates have $\Delta\beta_{k}<0$ and are accepted with exponentially small probability. 
As a result, the graph may remain unchanged for long periods before a rare downhill transition occurs. 
\begin{figure}[h]
    \begin{algorithm}[H]
        \caption{\textsc{MC search}}
        \label{Metropolis_algorith_format}
        \begin{flushleft}
            \textit{Input}: Random graph.\\
            \textit{Goal}: Maximize a chosen Betti number $\beta _{k}$.
        \end{flushleft}
        \begin{algorithmic}[1]
            \STATE \textbf{Step 1:} Initialize the graph $G$ and compute the initial Betti number $\beta(G)$.
            \STATE \textbf{Step 2:} Propose a random add/remove an edge to $G$ and update it to new graph $G'$.
            \STATE \textbf{Step 3:} Compute the change $\Delta\beta_{k}=\beta _{k}(G')-\beta _{k}(G)$ in the target Betti number ($\beta _{k}$).
            \STATE \textbf{Step 4:} Accept the new graph with probability 
                $p = \min\left(1, \re^{\Delta\beta_{k} / T}\right)$, where $T$ is the temperature.
            \STATE \textbf{Step 5:} If accepted, update $G \leftarrow G'$. Otherwise keep $G$ unchanged.
            \STATE \textbf{Step 6:} Repeat Steps 2–5 for a prescribed number of iterations.
            \begin{flushleft}
            \textit{Caveat}: Whenever $\beta_{0}\neq 1$, the protocol assigns the QFI the trivial value $\qfi=N$ by convention.
        \end{flushleft}
        \end{algorithmic}
    \end{algorithm}
\end{figure}
\subsubsection{Scaling of the QFI by using MC search}
\begin{figure}[h]
    \centering
    \includegraphics[width=\columnwidth]{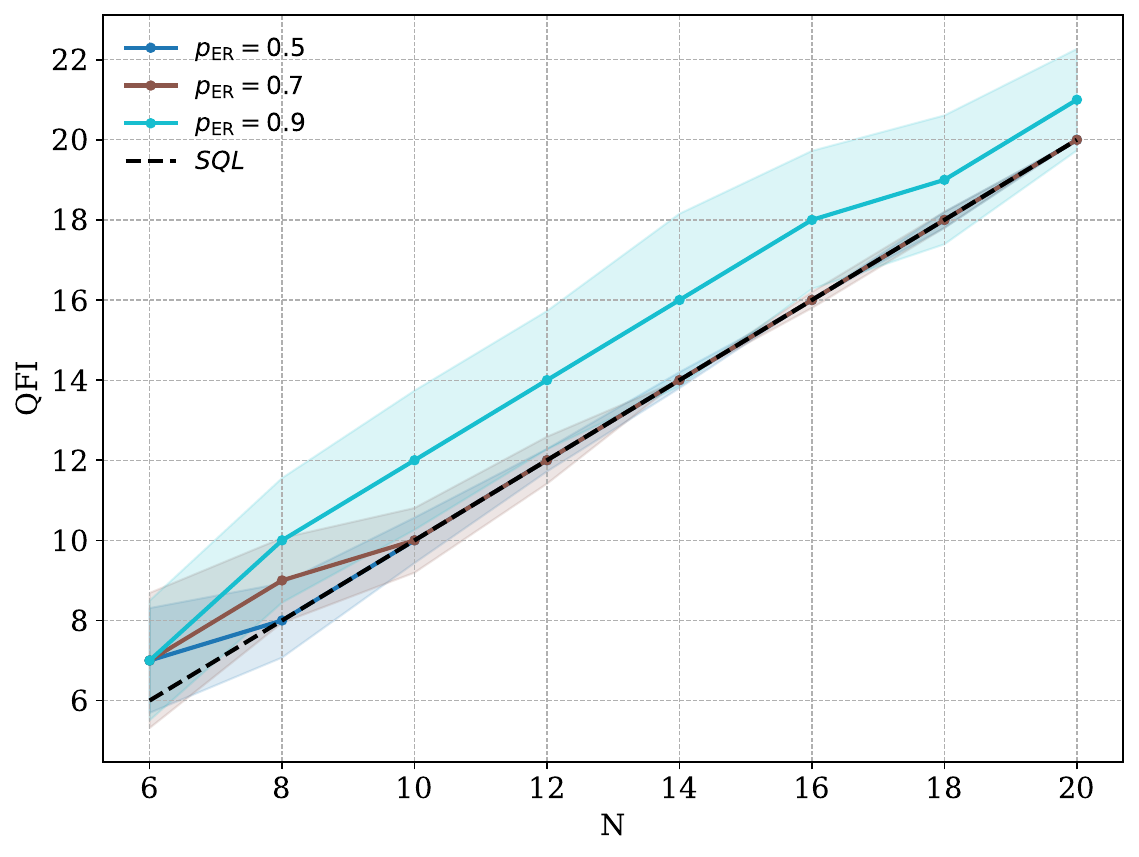}
    \caption{Mean QFI as a function of $N$, number of vertices in Erd\H{o}s--R\'enyi random graphs shown for different values of $p_{ER}$, averaged over $100$ realizations. The QFI exhibits linear scaling in $N$, consistent with SQL.}
    \label{QFI_random_graph_SQL}
\end{figure}

To assess whether the MC search can generate scaling of the QFI beyond the SQL, we first consider Erd\H{o}s--R\'enyi random graphs. 
Specifically, we compute the QFI for Erd\H{o}s--R\'enyi random graphs, $G(N,p_{ER})$, where $N$ is the number of vertices and each pair of distinct vertices is connected independently with probability $p_{ER}$. 
For two vertices $u,v$ in $G(N,p_{ER})$, since $N(u)=N(v)$ requires $u$ and $v$ to be non-adjacent, the probability that they have identical neighborhoods is
\begin{align}
    &\Pr[N(u)=N(v)]=\nonumber\\
    &\underbrace{(1-p_{ER})}_{u,v~\text{not-adjacent}}\underbrace{(1-2p_{ER}(1-p_{ER}))^{N-2}}_{\text{other}~(N-2)~\text{vertices}}.
\end{align}
Note that, otherwise, $v\in N(u)$ but $v\notin N(v)$. Hence, the expected number of pairs of vertices with identical neighborhoods is at most 
\begin{equation}
    N^{2}e^{-2p_{ER}(1-p_{ER})(N-2)}.
\end{equation}
Therefore, whenever $\log N/N \ll p_{ER} \leqslant 1/2$, with high probability every neighborhood-equivalence class has size $\nu_l=1$, and $\mathcal{Q}=N$ exactly. 
As shown in Fig.~\ref{QFI_random_graph_SQL}, the QFI exhibits a linear dependence on $N$ for different values of $p_{ER}$: for $p_{ER}=0.5$ the numerics match $\mathcal{Q}=N$ exactly, as predicted, while for larger $p_{ER}$ (e.g., $0.7$ and $0.9$) the scaling remains linear in $N$. 
This confirms that random graph ensembles do not provide any metrological advantage beyond the SQL.

\begin{figure}[h]
    \centering
    \includegraphics[width=\columnwidth]{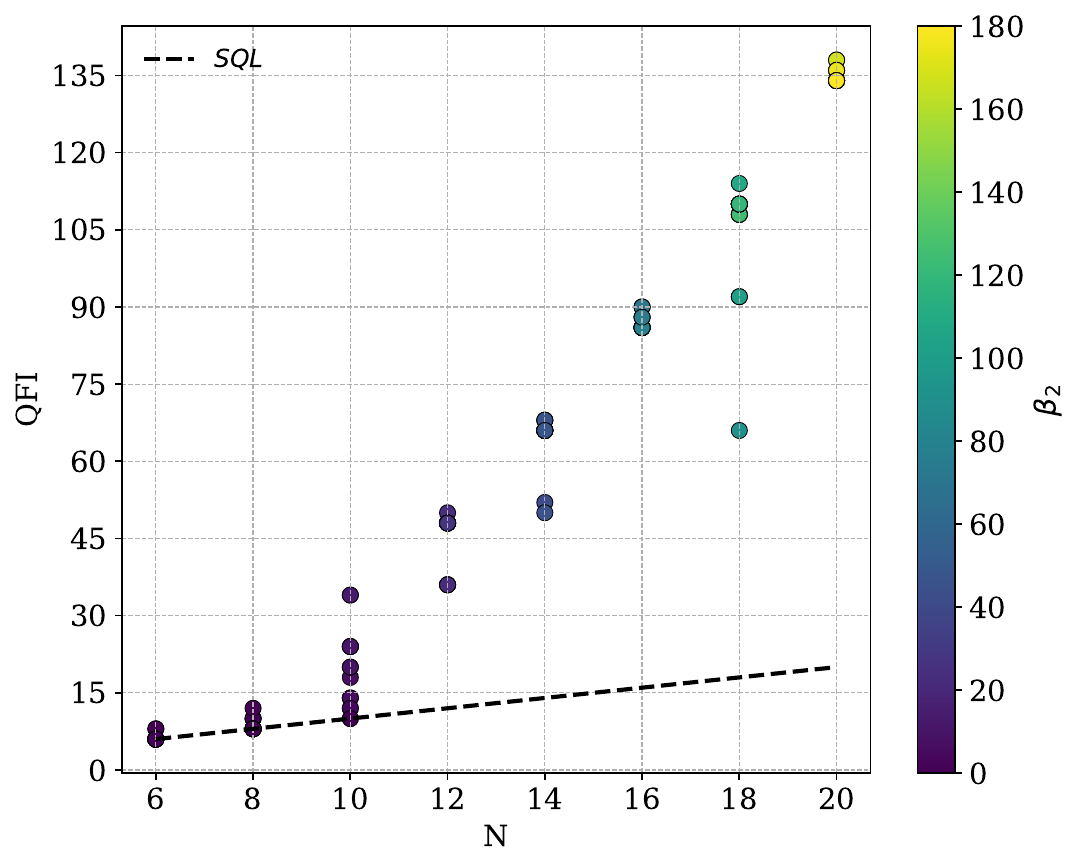}
    \caption{Final QFI values obtained for different system sizes $N$. Each point corresponds to an MC run with temperature $T=0.5$ and $35000$ edge-flip iterations, targeting the maximization of $\beta _{2}$. 
    For each $N$, the procedure is repeated $15$ times to account for stochastic variations in the MC search.}
    \label{Scaling_QFI_Metro}
\end{figure}

We next employ the MC search to maximize one of the Betti numbers, namely $\beta_{2}$ in this case. 
To examine how the QFI scales with the system size $N$ with temperature $T=0.5$, we run the MC search for a range of graph sizes, using $35000$ steps for each value of $N$. 
For every run, we record the final value of the QFI, and the resulting dependence of the QFI on $N$ is shown in Fig.~\ref{Scaling_QFI_Metro}. 
The data clearly indicate that the the MC search is successful and the QFI scales beyond the SQL. 
We repeat this experiment fifteen times to account for the stochasticity of the dynamics. 
As shown in Fig.~\ref{Scaling_QFI_Metro}, a subset of the realizations deviates from the overall trend. These realizations arise because the MC search becomes trapped in local maxima of the $\beta _{2}$ landscape and does not reach the global maximum within the allotted $35000$ iterations. 
All realizations are shown in Fig.~\ref{Scaling_QFI_Metro}, with the color indicating the corresponding value of $\beta _{2}$. 
Realizations exhibiting smaller values of $\beta _{2}$ typically correspond to lower values of the QFI, indicating that the degradation of the underlying topological structure is accompanied by a loss of metrological enhancement in the corresponding graph states. 
\subsubsection{Scaling of the QFI through MC search Betti optimization}
To further investigate the relationship between Betti numbers and the QFI in graphs, we initialize the MC search from an Erd\H{o}s--R\'enyi random graph $G(N=40,p_{ER}=0.05)$ and attempt to maximize a chosen Betti number using the MC search. 
Note that during the initial stage of the MC search, the proposed graph may have $\beta_{0}>1$. 
In this case, the QFI is assigned its trivial value $\qfi = N$. 
Once $\beta _{0}=1$, the QFI is evaluated according to \cref{QFI_graph_2}.
For each target Betti number, we run the MC search with temperature $T=0.5$ for $25000$ iterations, and repeat this procedure ten times to account for stochastic effects.

\begin{figure*} 
\includegraphics[width=1\linewidth]{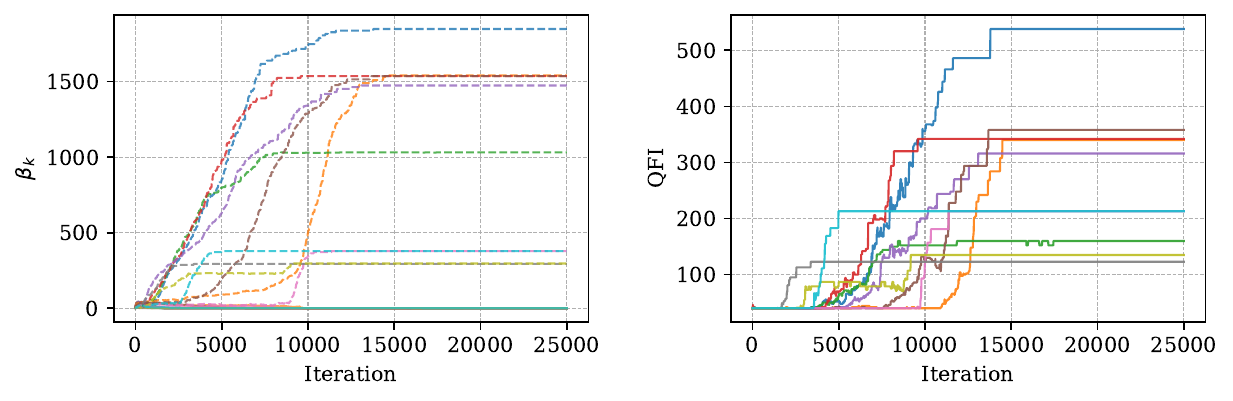}\quad\includegraphics[width=1\linewidth]{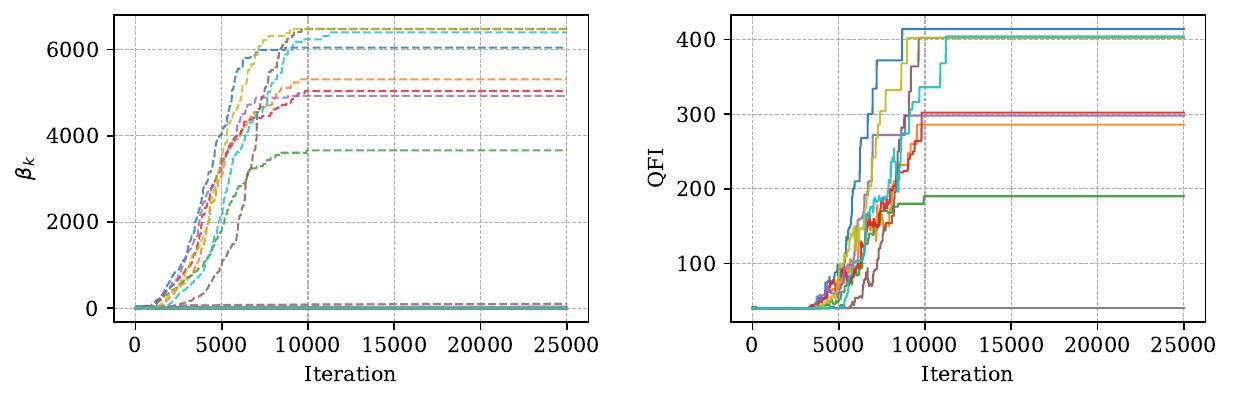}
\caption{Evolution of the Betti numbers and the QFI during the MC optimization. \emph{Top}: $\beta_{2}$; \emph{bottom}: $\beta_{3}$. Each run starts from an Erd\H{o}s--R\'enyi random graph with $N=40$, with $25000$ steps, and running the optimization process $10$ times.
\emph{Left panels}: Betti numbers as a function of MC iteration. 
\emph{Right panels}: QFI as a function of iteration. 
Dashed lines indicate the values of the nonzero Betti numbers ($\beta_{2}$ in the top panel and $\beta_{3}$ in the bottom panel) for different runs. All other Betti numbers remain zero (in the runs that converge) except $\beta_{0}=1$ (after the initial transient). At the beginning of the procedure, when $\beta_{0}\neq 1$ and isolated vertices are present, the QFI is set to $N$ by convention. Colours identify individual simulation runs, with the same colour corresponding to the same run in the left and right panels.
}
\label{QFI_Betti_Metro} 
\end{figure*}

Figure~\ref{QFI_Betti_Metro}--top shows the results obtained by maximizing $\beta_{2}$. 
During the MC search, the QFI grows and ultimately scales beyond the SQL. 
Inspection of the final graphs shows that they are bundled graphs with three islands. 
For $N=40$, the most symmetric bundled structure with three islands has sizes $\nu_1=\nu_{2}=13$, and $\nu_3=14$. 
For this configuration, the QFI reaches $534$, with the list of Betti numbers $(\beta_{0}=1,~\beta_{1}=0,~\beta_{2}=1872,~\beta_{3}=0)$. 
In Fig.~\ref{QFI_Betti_Metro}--top, the MC search finds a bundled graph with island sizes $\nu_1=15, \nu_{2}=13$, and $\nu_3=12$, for which the QFI reaches $538$ and the list of Betti numbers is $(\beta_{0}=1,~\beta_{1}=0,~\beta_{2}=1848,~\beta_{3}=0)$. 
This is the best result among the ten runs: one run reaches $\beta_2=1848$, while the others plateau at lower values, $1540$, $1474$, $1032$, and below. 
Thus, although the MC-searched graph has a smaller $\beta_{2}$ than the most symmetric configuration, it achieves a higher QFI. 
This observation suggests that the bundled structure provides both nontrivial topology and metrological enhancement, while the island sizes can affect the interplay between the maximum value of the corresponding Betti number and the QFI.

A similar behavior is observed when we instead maximize $\beta_{3}$, as shown in Fig.~\ref{QFI_Betti_Metro}--bottom. 
In this case, the MC search converges to graphs whose topology corresponds to a bundled graph with four islands. 
According to \cref{Betti_bundled_cyclic}, the maximum value of $\beta _{3}$ is $9^{4}=6561$, corresponding to four islands of equal size $m=10$, for which the QFI is $400$ (\cref{QFI_bundled_cyclic}). 
In Fig.~\ref{QFI_Betti_Metro}--bottom, the MC search finds a bundled graph with four islands of sizes $\nu_1=9$, $\nu_2=8$, $\nu_3=10$, and $\nu_4=13$, yielding a QFI of $414$ and the associated Betti vector $(\beta_0=1,\,\beta_1=0,\,\beta_2=0,\,\beta_3=6048)$. It also identifies a bundled graph with four islands of sizes $\nu_1=9$, $\nu_2=9$, $\nu_3=11$, and $\nu_4=11$, yielding a QFI of $404$ and the Betti vector $(\beta_0=1,\,\beta_1=0,\,\beta_2=0,\,\beta_3=6400)$.

Two other runs reach a larger value of $\beta_3=6480$, corresponding to island sizes $\nu_1=9$, $\nu_{2}=10$, $\nu_3=10$, and $\nu_4=11$, but yield a slightly smaller QFI of $402$. 
Thus, the run with $\beta_3=6048$ achieves the highest QFI among these cases, illustrating the interplay between the maximum value of the corresponding Betti number and the QFI. 
Again, we find that maximizing one of the Betti numbers yields a graph with nontrivial topology and enhanced QFI scaling, while there is an interplay between the maximum value of the corresponding Betti number and the QFI, which can be tuned through the number of nodes in each island.

Taken together, these results demonstrate that starting from a random graph and maximizing the $k$-th Betti number drives the graph toward nontrivial topologies whose corresponding graph states exhibit metrological enhancement beyond the SQL. 
This behavior is consistent with the known extremal property of complete multipartite graphs, which maximize the corresponding Betti number~\cite{kozlov_convex_1997,beers2025extremal} and whose graph states exhibit QFI beyond the SQL~\cite{shettell_graph_2020}. 
Furthermore, it demonstrates the stability of the relationship beyond extremal settings.
More importantly, Fig.~\ref{QFI_Betti_Metro} shows that the QFI grows with $\beta_k$ once the bundled structure forms. 
In particular, although the QFI remains approximately constant at $N=40$ during the initial stage of the MC search, where $\beta_k$ increases, it starts to grow as the bundled structure develops. 
In the runs that converge to the bundled graph, the QFI reaches a value beyond the SQL, indicating that the metrological enhancement is associated with the formation of the corresponding topological structure.

\section{Conclusion}\label{Sec_conclusion}
In this work, we have investigated the relationship between the QFI of graph states and the topological properties of their underlying graphs, as characterized by Betti numbers. 
Through numerical analysis, we found numerical evidence for a connection between the metrological performance of graph states, quantified by the scaling of the QFI with system size, and the topology of the associated graphs.

In the graph ensembles studied here, the largest QFI values are attained by bundled graphs, whose Betti numbers vanish except for $\beta_0$ and a single higher-order Betti number $\beta_k$. Under weak edge deletion, several higher-order Betti numbers become nonzero, while the QFI exhibits SQL scaling. Under edge addition, the QFI and the higher-order Betti numbers decay together. Bundled graphs with $l$ islands have $\qfi=N^2/l$, as given by Eq.~\eqref{QFI_bundled_cyclic}, whereas deviations from this structure lead to a crossover toward $\mathcal{Q}\approx N$.

We further studied an optimization protocol in which, starting from random graphs with a fixed number of qubits, a selected Betti number is iteratively maximized. 
This procedure typically leads to graph configurations with enhanced metrological performance. 
At convergence, the graph develops a bundled structure with nearly maximal values of the targeted Betti number, while the remaining Betti numbers are suppressed. 
Moreover, once the bundled structure forms, the QFI grows with the targeted Betti number along
each trajectory, and in the runs that converge to
a bundled graph, the resulting QFI lies in the super-linear regime. 
Note that optimizing the QFI directly leads to a qualitatively different outcome from the topological optimization considered here. 
Maximizing \cref{QFI_graph_2} yields a star graph, where a central node connects to all other nodes, while those nodes are not connected to each other (since a vertex cannot be its own neighbor, the largest possible class has size $(N-1)$, which maximizes the corresponding sum of squares). 
Hence, the QFI for the star graph is $\qfi =1+(N-1)^{2}$. 
However, the star graph is topologically trivial, with vanishing higher-order Betti numbers $(\beta_{0}=1,~\beta_{1}=0,~\beta_{2}=0,~\beta_{3}=0,\cdots)$. 
Therefore, maximizing the QFI alone drives the graph toward a highly asymmetric but topologically simple structure, whereas maximizing higher-order Betti numbers produces graphs with nontrivial topology and enhanced metrological performance. 
This contrast highlights that the topological optimization explored here provides a distinct route to generating metrologically useful graph states.

Our results provide evidence that the QFI encodes not only geometric information about quantum states, but also, in the graph ensembles studied here, signatures of the underlying topological structure of the associated graphs.

More broadly, these findings suggest a connection between information-theoretic aspects of quantum metrology and algebraic topology, highlighting topological invariants as potential indicators of metrological performance in multipartite quantum systems. 
This work opens several directions for future research, in particular the development of an analytic framework clarifying the relationship between QFI scaling and Betti numbers. 
Such an understanding could reveal the topological origin of quantum-enhanced metrological scaling and guide the systematic design of highly sensitive quantum states.

\textit{Code and data statement}--- Code and data are available from the authors upon reasonable request. 

\textit{Acknowledgements}--- We thank Alice Barthe, Xiaoyu Liu, and Katerina Saiti for insightful discussions and numerical support. 
M.H. \& J.T. acknowledge the support received from the European Union’s Horizon Europe research and innovation programme through the ERC StG FINE-TEA-SQUAD (Grant No. 101040729). 
P.E.~acknowledges the support received through the NWO-Quantum Technology program (Grant No.~NGF.1623.23.006, DOI:~\url{https://doi.org/10.61686/RBLDP65717}) and funding by the Carl-Zeiss-Stiftung (CZS Center QPhoton). 
Parts of this work were performed by using the compute resources from the Academic Leiden Interdisciplinary Cluster Environment (ALICE) provided by Leiden University.

Views and opinions expressed are, however, those of the author(s) only and do not necessarily reflect those of the funding institutions. Neither of the funding institutions can be held responsible for them.
\bibliography{references}
\end{document}